\documentclass{article}
\usepackage{waspaa17,amsmath,graphicx,url,times}
\usepackage{color}

\title{Frequency Domain Singular Value Decomposition for Efficient Spatial Audio Coding}

\name{Sina Zamani, Tejaswi Nanjundaswamy, Kenneth Rose  }
\address{Department of Electrical and Computer Engineering, University of California Santa Barbara, CA 93106\\ 
E-mail: \{sinazmn, tejaswi, rose\}@ece.ucsb.edu} 

\begin{document}

\maketitle

\begin{sloppy}

\begin{abstract}
Advances in virtual reality have generated substantial interest in accurately reproducing and storing spatial audio in the higher order ambisonics (HOA) representation, given its rendering flexibility. Recent standardization for HOA compression adopted a framework wherein HOA data are decomposed into principal components that are then encoded by standard audio coding, i.e., frequency domain quantization and entropy coding to exploit psychoacoustic redundancy. A noted shortcoming of this approach is the occasional mismatch in principal components across blocks, and the resulting suboptimal transitions in the data fed to the audio coder. Instead, we propose a framework where singular value decomposition (SVD) is performed after transformation to the frequency domain via the modified discrete cosine transform (MDCT). This framework not only ensures smooth transition across blocks, but also enables frequency dependent SVD for better energy compaction. Moreover, we introduce a novel noise substitution technique to compensate for suppressed ambient energy in discarded higher order ambisonics channels, which significantly enhances the perceptual quality of the reconstructed HOA signal. Objective and subjective evaluation results provide evidence for the effectiveness of the proposed framework in terms of both higher compression gains and better perceptual quality, compared to existing methods.

\end{abstract}


\section{Introduction}
\label{sec:intro}

The ambisonics paradigm was originally developed \cite{HOA1, HOA2} as a promising technique for reproducing three dimensional sound fields. However, it was not commercially successful at the time due to poor directionality and the limited size of the ``sweet spot''. Later, with the advent of higher order ambisonics (HOA)\cite{HOA3}, the approach was extended by sound field decomposition into higher order spherical components, resulting in improved localization, spatial resolution, and increased size of the sweet spot. Recent advances in virtual reality and many related applications such as streaming of real time music performances or cinematic scenes, have fueled interest in HOA to accurately reproduce spatial audio. HOA data (typically obtained from a microphone array) are high dimensional and pose significant challenges on storage and transmission for practical applications, which motivate the development of novel effective compression techniques to considerably reduce the required bit-rate.

Early approaches \cite{burnett2008encoding} directly encoded individual HOA channels with Advanced Audio Coding (AAC), independent of other channels. It was observed that allocating more bits to lower order components increases the sound quality in the sweet spot but decreases spatial resolution. As such approaches clearly neglect inter-channel redundancies, later work \cite{hellerud2009spatial} employed a lossless compression technique to exploit inter-channel correlations, by choosing one of the channels as reference for predicting the other channels, followed by encoding of the prediction residue.

Recently, a new spatial audio coding standard, MPEG-H 3D Audio \cite{herre2015mpeg}, has emerged. The HOA input is decomposed into predominant sound elements and ambient background components, using standard singular value decomposition (SVD), and each of these are coded separately via an AAC based coder, where quantization and entropy coding are performed in the frequency domain to exploit psychoacoustic redundancies. While good broadcast quality has been reported for bit-rates around 300 kbps \cite{peters2015scene}, the premise of this paper is that higher compression efficiency and better perceptual quality can be achieved by employing SVD in the frequency domain. In the MPEG-H approach, there is often a mismatch of principal components across blocks, both in terms of order of components and their respective basis vectors. MPEG-H employs an elaborate matching technique in combination with an overlap-add technique to mitigate this shortcoming. However, transitions between blocks remain suboptimal and introduce inefficiencies in the core codec and degrade the perceptual quality. Our approach completely eliminates this issue as we first transform to frequency domain via MDCT which ensures smooth transition across blocks with its built-in overlap.
Moreover, optimal SVD can now be adapted to different frequencies, instead of a compromise decomposition for the entire spectrum. Finally, we employ noise substitution in a novel way to compensate for ambient energy loss and further improve perceptual quality of the rendered HOA data.

\section{MPEG-H approach for compression of HOA data}
\label{sec:MPEG}

\begin{figure}[t]
  \centering
  \centerline{\includegraphics[width=1.05\columnwidth]{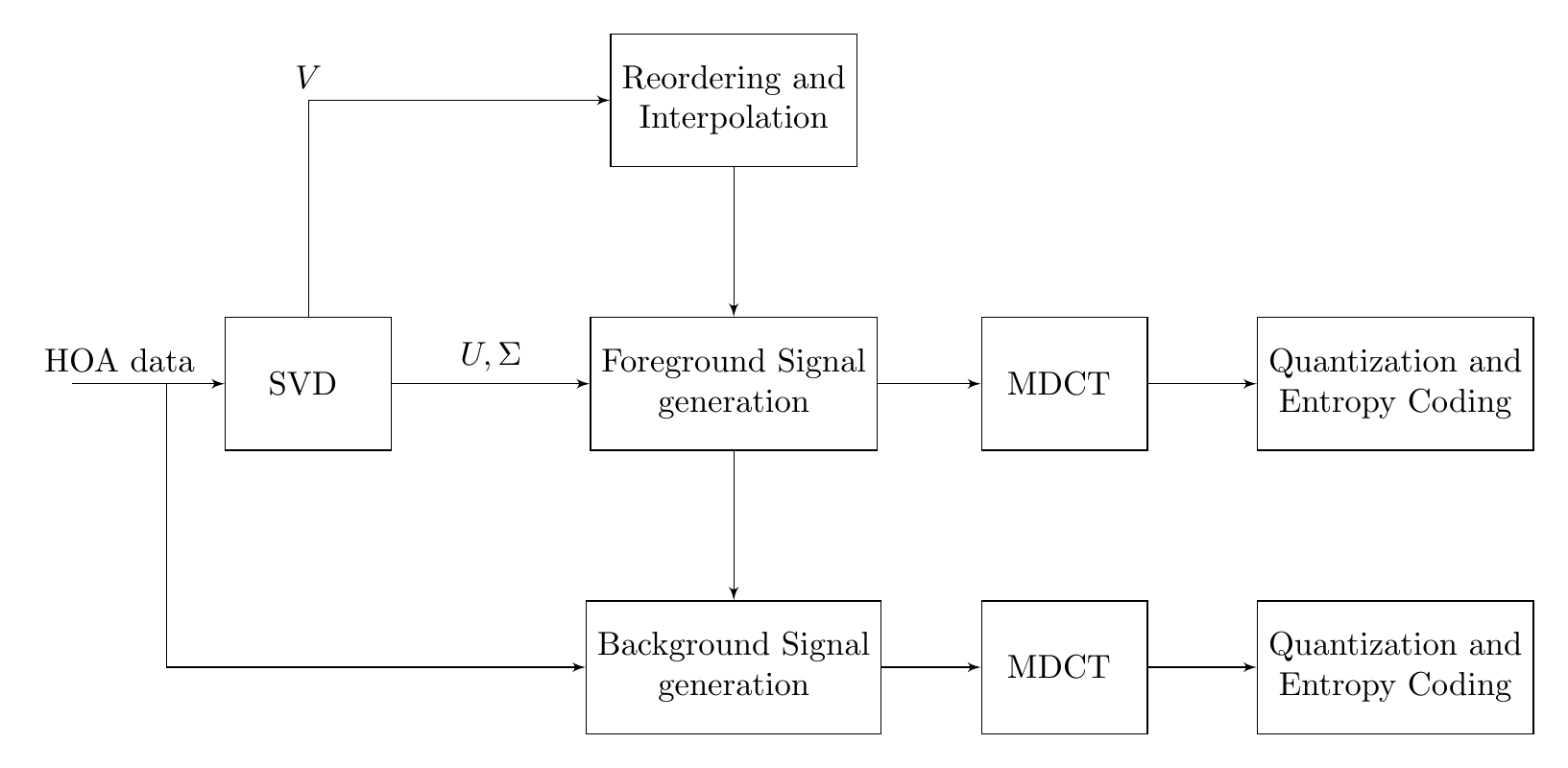}}
  \vspace{-12pt}
\caption{Overview of MPEG-H encoder}
\label{fig:mpeghenc}
\vspace{-15pt}
\end{figure}

The MPEG HOA encoder \cite{mpeg_3d} processes the input HOA data over frames of length $2L$ ($L=1024$) with 50\% overlap. Let the number of HOA channels be $M = (N+1)^2$, where $N$ is the ambisonics order. The encoder operates on the current frame data $X_f$, which is an $2L \times M$ matrix, and performs standard singular value decomposition (SVD),
\begin{equation}
X_f = U_f {\Sigma}_f V_f^T, 
\label{eqn:mpegdecom}
\end{equation}
where $U_f$ is an $2L \times 2L$ unitary matrix, ${\Sigma}_f$ is a $2L \times M$  rectangular diagonal matrix with non-zero elements on the diagonal and $V_f$ is an $M\times M$ unitary matrix. The SVD construction ensures that predominant components, corresponding to the largest $r$ singular values have as basis vectors the first $r$ columns of $V_f$. Let $V_f$ be truncated to the first $r$ columns, and further be independently or deferentially quantized to ${\hat{V}}_f$ and sent to the decoder as side information for each frame, so as to enable it to transform back the predominant components to the ambisonics domain. To keep encoder and decoder in sync, the quantized ${\hat{V}}_f$ is used to generate the predominant components $\tilde{Y}_{f}$ (now an approximation of first $r$ columns of $U_f {\Sigma}_f$), as,
\begin{equation}
\tilde{Y}_{f} = {{X}}_f{{\hat{V}}_f} {({\hat{V}}_f^{T}{{\hat{V}}_f})}^{-1}.
\label{eqn:mpegupd}
\end{equation} 
Note that the inverse term is for renormalization of the quantized basis vectors (to maintain unitarity). The next step is to code the predominant or foreground components, each corresponding to a column of $\tilde{Y}_f$, using separate instances of the core audio codec. This requires concatenating components across frames. However, 
since SVD arranges the basis vectors based on the singular value magnitudes, the same foreground component might change position in $\tilde{Y}$ from frame to frame depending on the magnitude of its singular value relative to others. This can result in noticeable blocking artifacts if blindly concatenated foreground components are fed to the core codec.
While there are several approaches to reorder and match components with the previous frame, we employ the magnitude of correlation between column vectors of $\hat{V}_f$ and $\hat{V}_{f-1}$ as the criterion in an Hungarian matching algorithm \cite{kuhn1955hungarian}, which we found to be effective.

Even with matched components, simple concatenation across frames would introduce noticeable artifacts as a small change in the basis vector causes some mismatch at the frame boundary. Hence the encoder interpolates the column vectors of $\hat{V}$ between current frame and previous frame to ensure continuity over time. Specifically, a different transform matrix is used for each sample of the current frame, whose column vectors are obtained as,
\begin{equation}
\begin{aligned}
\bar{v}^i_f(l) = (1-w(l)) {\hat{v}_{f-1}^i}   +  w(l) {\hat{v}_{f}^i}, \\ l=0 \text{ to } L-1,~ i=0 \text{ to } r-1,
\vspace{-3pt}
\label{eqn:interpolation}
\end{aligned}
\end{equation} 

where ${\hat{v}_{f}^i}, {\hat{v}_{f-1}^i}$ are the $i$th matched column vectors of $\tilde{V}$ for current and previous frames, $\bar{v}^i_f(l)$ is $i$th column vector for sample $l$ in current frame and $w(l)$ is a window function, which may be the triangular or Hanning window. The interpolation should also account for the fact that the vectors might get negated from one frame to next frame by performing a sign correction when needed. 

An approximation of the HOA data, $\tilde{X}_f$, is generated by transforming the foreground components back to the ambisonics domain, which is then subtracted from the original data to produce the ambient (or background) HOA data. 
The foreground components are coded using separate instances of the core audio codec. The order of background HOA data is then reduced (from $N$ to some $t$) and this lower order HOA data are also coded using the core audio codec. An illustration of the MPEG-H approach is shown in Figure \ref{fig:mpeghenc}.


\section{Frequency Domain SVD for HOA data compression}
\label{sec:geneed}
\vspace{-0.5mm}
\begin{figure}[t]
  \centering
  \centerline{\includegraphics[width=1.15\columnwidth]{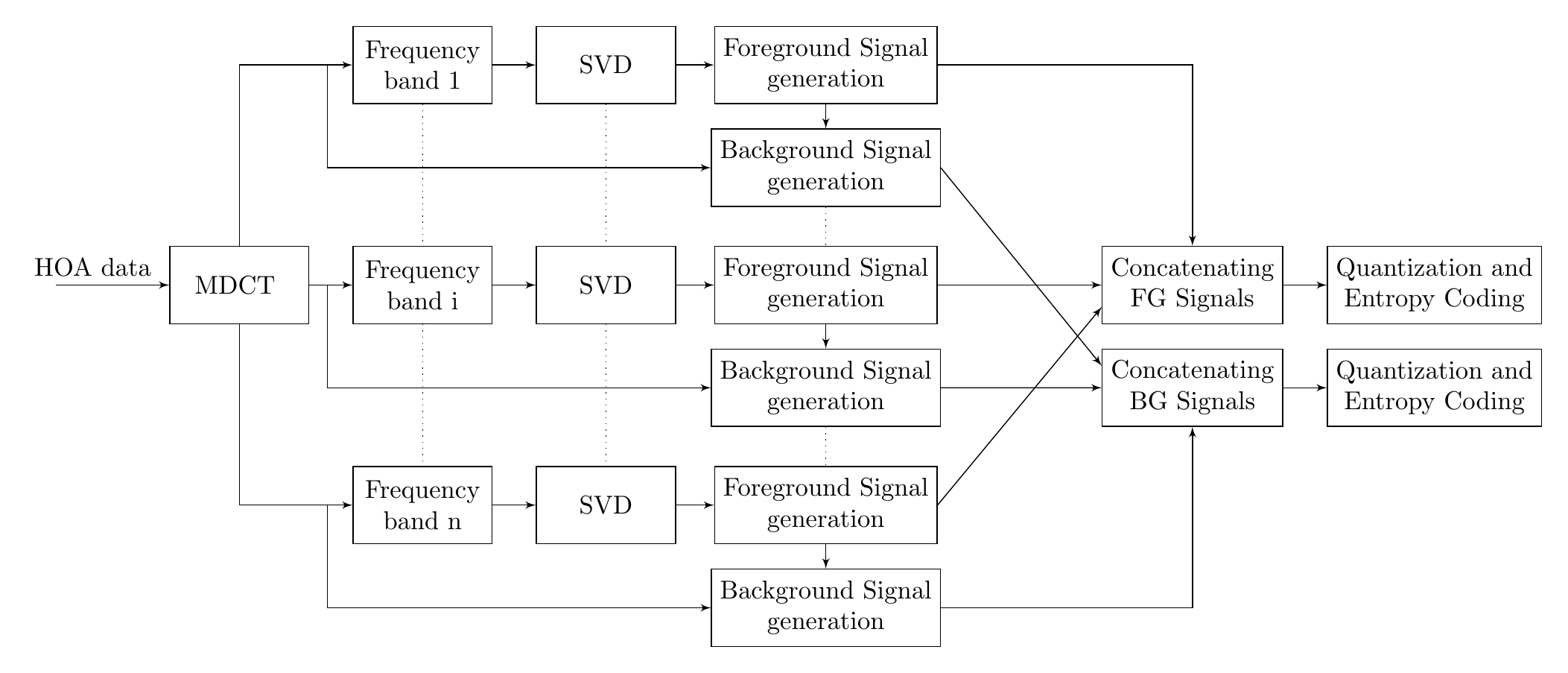}}
\vspace{-12pt}
\caption{Overview of proposed method encoder}
\vspace{-12pt}
\label{fig:propenc}
\end{figure}
\vspace{-0.5mm}
Clearly, the MPEG-H approach performs an elaborate process of matching and interpolating transform basis vectors of consecutive frames to improve their continuity over time and to mitigate the artifacts stemming from blockwise SVD application. We propose to circumvent this underlying and fundamental shortcoming with a framework wherein SVD is employed {\em after} transformation to frequency domain via MDCT, which naturally achieves the required smoothness with its built-in overlap. Moreover, this framework enables the significant flexibility to make both the SVD and the number of components to be retained, adaptive to frequency, instead of using a compromise for the varying needs of different frequency bands.


In the proposed approach, the HOA data are processed in the encoder after segmenting each HOA channel into $50\%$ overlapped frames of length $2L$. The samples of each channel are separately transformed via MDCT after windowing to obtain the transformed data for the current frame, $S_f$, which is an $L \times M$ matrix. $S_f$ is now divided into different frequency bands, $ S^T_f = [ S^T_{f_1} S^T_{f_2} ... \; S^T_{f_n} ]$ where $n$ is the number of frequency bands with lengths $l_1,l_2,...,l_n$ and $\sum_i l_i=L$. For each frequency band, a different SVD is obtained, $S_{f_i} =  U_{f_i} {\Sigma}_{f_i} V_{f_i}^T $, the top $r_i$ components are retained (which may vary over bands), and the correspondingly truncated $V_{f_i}$ are coded to $\hat{V}_{f_i}$ and sent to the decoder as side information. The prominent components, $\tilde{Y}_{f_i}$, are now obtained similar to \eqref{eqn:mpegupd} for each band and concatenated together to form $\tilde{Y}^T_f = [ \tilde{Y}^T_{f_1} \tilde{Y}^T_{f_2} ... \; \tilde{Y}^T_{f_n} ]$. Given the foreground components, an approximate $\tilde{S }_{f}$ is obtained and 
subtracted from original data $S_{f}$ to obtain the residual. The columns of  $\tilde{Y}_f$ are then separately quantized and entropy coded similar to the AAC codec. Finally, the residual is reduced in ambisonics order, and quantized and entropy coded similar to the AAC codec. An illustration of the proposed approach is shown in Figure \ref{fig:propenc}.
\vspace{-7pt}
\subsection{Side Information Compression}
\label{sec:sidei}

To exploit the temporal correlations between transform matrices of consecutive frames, the Hungarian algorithm \cite{kuhn1955hungarian} is employed to match the column vectors of $V_{f_i}$ matrices of consecutive frames corresponding to $i$th frequency band based on correlation coefficients. We used a scalar prediction coefficient (equal to correlation coefficient) for each vector. We selected approximately 10,000 frames from third order ambisonics files as training set to design a quantizer for prediction coefficients and prediction residuals using Generalized Lloyd Algorithm (GLA). 
\vspace{-7pt}
\subsection{Noise Substitution}
\label{sec:noise}

Reduction of ambient HOA data order results in suppression of ambient energy.
The MPEG-H approach compensates for this lost energy by determining and applying gains to the order reduced ambient components. 
However, we observed in our experiments that this approach did not offer sufficient improvement of perceptual quality, 
while increasing the bit-rate. 

We found that certain background sounds can only be reproduced if all the channels of ambient components are coded using the core codec. Since this may often represent a prohibitive cost in bit-rate, we propose an approach where along with encoding reduced order ambient data with the core codec, we substitute the discarded channels with perceptual noise in frequency bins across these channels. Specifically, we measure the spectral flatness in each of the discarded channels for each of the 49 AAC frequency groups as following,
\vspace{-12pt}
\begin{equation}
\text{Flatness}_f^{ij} = \frac{exp( \frac{1}{|{B}_f^{ij}|} \sum_{ k \in {B}_f^{ij} } \ln{{B}_f^{ij}[k]}   )}{\frac{1}{|{B}_f^{ij}|} \sum_{ k \in {B}_f^{ij} } {B}_f^{ij}[k]} , 
\label{eqn:mpegprom}
\end{equation}	

where ${B}_f^{ij}$ are the power spectrum coefficients for channel $i$ and frequency group $j$ of the current frame background data. This flatness is averaged across all channels for each frequency group and if the average is larger than a threshold, randomly generated noise is added to these frequencies in the discarded ambient channels at the decoder. The average energy across all channels in this frequency group is sent to the decoder to generate the noise at the original content energy.
Thus, a maximum of 49 energy values are coded and sent to the decoder for each frame.

\vspace{-12pt}
 \begin{table*}[t]
\centering
\begin{tabular}{cccc}
\hline
& \multicolumn{2}{c}{Operating point } \\
\hline
Sequence      & \begin{tabular}[c]{@{}c@{}}  around 308 kbps \end{tabular} & \begin{tabular}[c]{@{}c@{}} around 375 kbps \end{tabular}
 & \begin{tabular}[c]{@{}c@{}} around 500 kbps \end{tabular}
\\ \hline
\textit{2src\_conv\_office\_toa}       & 7.83\% & 8.03\%  & 8.37\% \\ \hline
\textit{A Round Around-Eigen-TOA-ACN}          &    -0.8\%  & 0\%  & 1.41\%  \\ \hline
\textit{doll\_intro\_3oa}      &    6.64\%   & 6.93\%  & 7.26\%  \\ \hline
\textit{helicopter\_fountain\_toa}         &  4.58\% & 5.96\%  & 7.02\%  \\ \hline

\textit{lyon\_toa}       &   3.98\% & 4.72\%  & 3.58\%  \\ \hline

\textit{Murmur2\_toa}          &   6.50\% & 8.89\%  & 10.9\%  \\ \hline\hline
Average         & 4.79\% & 5.75\%  & 6.44\%  \\ \hline
\end{tabular}
\vspace{-8pt}
\caption{Proposed framework's reduction in bit-rate}
\vspace{-4.5mm}
\label{tbl:obj}
\end{table*}

\section{Experimental Results}
\label{sec:results}

To validate the efficacy of the proposed approach we conducted objective and subjective experiments. The experiment was on a dataset of recordings provided by Google, which consist of 6 third order ambisonics files.  
As the software for MPEG-H encoder is not yet publicly available, we implemented our own representative version of it, as described in Section \ref{sec:MPEG}, based on the published patents \cite{patent, patent2} and the standard documentation \cite{mpeg_3d} which serves as a baseline for comparison. Other than the explicit contributions of the new approach, the competitors are identical in terms of options enabled, etc. All side information is accounted for in the total bit-rate. 

In all the experiments $r=r_i=4, \forall i$ and $t= 1$, that is, the number of foreground and background channels are both set to 4 for all frames, which results in a total of 8 components being encoded with the core codec. In the proposed approach, we divided the frequency data into 4 uniformly sized bands and a different transform is obtained for each frequency band. While employing frequency dependent SVD always results in better compaction of energy, this does not always translate to improved RD performance for the fixed quantizers and entropy coders employed. We believe this limitation can be addressed by redesigning the quantizers and entropy coders for the new statistics. In order to obtain preliminary results we employed the ``shortcut'' of providing two encoding modes per frame, of using a single frequency band (mode $m_f = 0$), or using 4 frequency bands (mode $m_f=1$), and selecting the one which minimizes the RD cost. When the mode switches between frames, the transform matrix (or matrices) of current frame are predicted from the best available previous transform matrix (or matrices). 




\subsection{Objective Results}
\label{sec:obj}

Note that perceptual distortion optimization for foreground data obtained through SVD, especially in comparison to background data in ambisonics domain, is still an open problem. 
To obtain preliminary objective results, we simply encoded both the competing methods to minimize the bit-rates for a given maximum quantization noise to mask ratio (MNMR) constraint for all bands of all channels. Investigation of the true objective perceptual distortion measure and its corresponding optimization approach is part of future work.
Percentage reduction in bit-rate for the proposed method in comparison to the MPEG-H approach, obtained at different operating points is presented in Table~\ref{tbl:obj}. Clearly, there is a consistent improvement in performance for the proposed framework. 




\vspace{-5pt}
\subsection{Subjective Results}
\label{sec:subj}

We conducted subjective evaluations to determine the true perceptual gains using the MUSHRA listening tests \cite{mushra}. This is particularly important given the above reservations about the ability of the objective measure to fully capture the perceptual quality. The test items were scored on a scale of 0 (bad) to 100 (excellent) and the tests were conducted with 8 listeners. We extracted 10s portions of each file for evaluation. The test files includes challenging scenes with speech, music and objects moving. A binaural renderer was deployed to convert the reconstructed HOA coefficients to stereo signals. Randomly ordered 4 versions of each audio sample (including a hidden reference, a 3.5 kHz low-pass filtered anchor, the encoded file using the proposed method and the encoded file using the MPEG method) were presented to the listeners. For these tests, the bit-rate were matched for each competing file. The subjective evaluation results, including the mean and 95\% confidence intervals, as presented in Figure \ref{fig:mushra} clearly demonstrates the substantially improved quality. This margin of improvement could not have been predicted from the moderate gains observed in objective results, clearly highlighting the critical need for further research in developing an appropriate objective perceptual distortion measure and corresponding optimization approach. The files used for these subjective tests are shared in \cite{subj_link}.

\begin{figure}[t]
  \centering
  \centerline{\includegraphics[width=0.95\columnwidth]{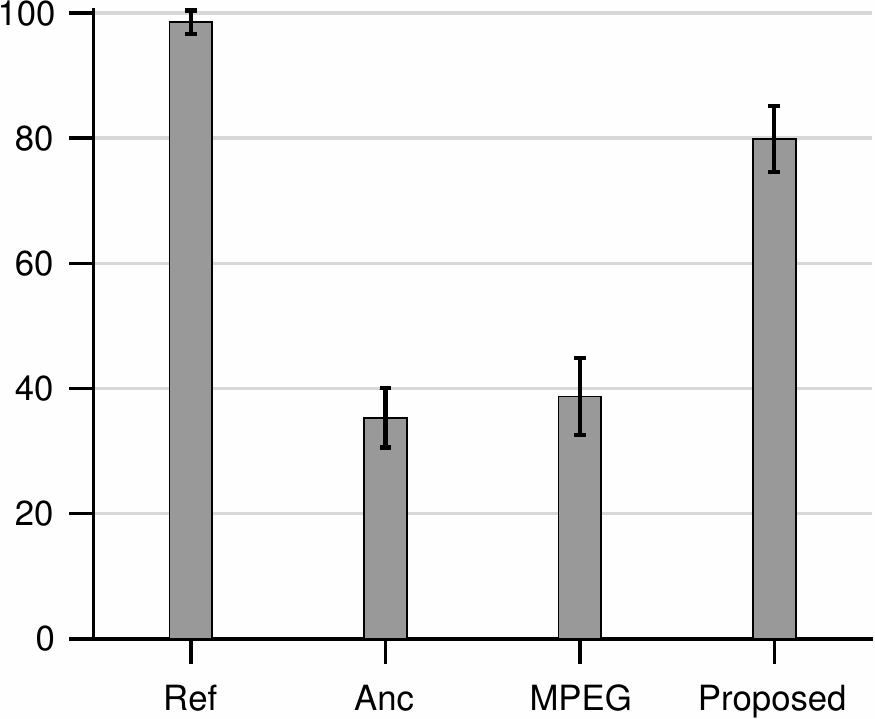}}
  \vspace{-7pt}
\caption{MUSHRA listening test results comparing the encoding techniques}
\vspace{-4.5mm}
\vspace{-5pt}
\label{fig:mushra}
\end{figure}

\section{Conclusion}
\label{sec:concl}

This paper presents a new framework for compression of higher order ambisonics data by first transforming the coefficients to MDCT domain and then decomposing into principal components. Unlike the current approaches, which suffer from suboptimal transitions between frames, the proposed approach not only ensures smooth transitions, it also enables frequency dependent decomposition and selection of dominant components. Furthermore, a novel way of employing noise substitution is introduced to enhance the perceptual quality of final reconstructions. Objective and subjective results illustrate the effectiveness of proposed approach with significant performance improvements. Future research includes optimally deciding number and size of frequency bands, frequency dependent optimization of the number of foreground and background components, redesign of quantizers and entropy coders, and investigation of better objective perceptual distortion measure.

\section{Acknowledgment}
\label{sec:concl}

This research was supported by Google, Inc. We are particularly grateful to Jan Skoglund and Drew Allen from Google for useful discussions and providing us with the Ambisonics dataset and the binaural renderer used in Sec.~\ref{sec:results}.

\bibliographystyle{IEEEtran}
\bibliography{refs17}

\begin{thebibliography}{10}
\providecommand{\url}[1]{#1}
\def\UrlFont{\rmfamily}
\providecommand{\newblock}{\relax}
\providecommand{\bibinfo}[2]{#2}
\providecommand\BIBentrySTDinterwordspacing{\spaceskip=0pt\relax}
\providecommand\BIBentryALTinterwordstretchfactor{4}
\providecommand\BIBentryALTinterwordspacing{\spaceskip=\fontdimen2\font plus
\BIBentryALTinterwordstretchfactor\fontdimen3\font minus
  \fontdimen4\font\relax}
\providecommand\BIBforeignlanguage[2]{{%
\expandafter\ifx\csname l@#1\endcsname\relax
\typeout{** WARNING: IEEEtran.bst: No hyphenation pattern has been}%
\typeout{** loaded for the language `#1'. Using the pattern for}%
\typeout{** the default language instead.}%
\else
\language=\csname l@#1\endcsname
\fi
#2}}

\bibitem{HOA1}
M.~A. Gerzon, ``Periphony: With-height sound reproduction,'' \emph{Journal of
  the Audio Engineering Society}, vol.~21, no.~1, pp. 2--10, 1973.

\bibitem{HOA2}
M.~Gerzon, ``Ambisonics in multichannel broadcasting and video,'' \emph{Journal
  of the Audio Engineering Society}, vol.~33, no.~11, pp. 859--871, 1985.

\bibitem{HOA3}
J.~Daniel, S.~Moreau, and R.~Nicol, ``Further investigations of high-order
  ambisonics and wavefield synthesis for holophonic sound imaging,'' in
  \emph{Audio Engineering Society Convention 114}, 2003.

\bibitem{burnett2008encoding}
E.~Hellerud, I.~Burnett, A.~Solvang, and U.~P. Svensson, ``Encoding higher
  order ambisonics with {AAC},'' in \emph{Audio Engineering Society Convention
  124}, 2008.

\bibitem{hellerud2009spatial}
E.~Hellerud, A.~Solvang, and U.~P. Svensson, ``Spatial redundancy in higher
  order ambisonics and its use for lowdelay lossless compression,'' \emph{IEEE
  International Conference on Acoustics, Speech, and Signal Processing
  (ICASSP)}, pp. 269--272, 2009.

\bibitem{herre2015mpeg}
J.~Herre, J.~Hilpert, A.~Kuntz, and J.~Plogsties, ``{MPEG-H 3D} audio—the new
  standard for coding of immersive spatial audio,'' \emph{IEEE Journal of
  Selected Topics in Signal Processing}, vol.~9, no.~5, pp. 770--779, 2015.

\bibitem{peters2015scene}
N.~Peters, D.~Sen, M.-Y. Kim, O.~Wuebbolt, and S.~M. Weiss, ``Scene-based audio
  implemented with higher order ambisonics {(HOA)},'' in \emph{SMPTE Annual
  Technical Conference and Exhibition}, 2015, pp. 1--13.

\bibitem{mpeg_3d}
\emph{{Information technology -- High efficiency coding and media delivery in
  heterogeneous environments -- Part 3: 3D audio}}, ISO/IEC Std. ISO/IEC
  JTC1/SC29 23\,008-3:2015, 2015.

\bibitem{kuhn1955hungarian}
H.~W. Kuhn, ``The hungarian method for the assignment problem,'' \emph{Naval
  research logistics quarterly}, vol.~2, no. 1-2, pp. 83--97, 1955.

\bibitem{patent}
D.~Sen and N.~Peters, ``Interpolation for decomposed representations of a sound
  field,'' Dec.~4 2014, {WO2014194099 A1}.

\bibitem{patent2}
D.~Sen and S.-U. Ryu, ``Compression of decomposed representations of a sound
  field,'' Dec.~4 2014, {US20140358563 A1}.

\bibitem{mushra}
\emph{{Method of Subjective Assessment of Intermediate Quality Level of Coding
  Systems}}, ITU Std. {ITU-R Recommendation, BS 1534-1}, 2001.

\bibitem{subj_link}
\BIBentryALTinterwordspacing
``{HOA} subjective listening test files,'' (Date last accessed 28-Apr-2017).
  [Online]. Available: \url{https://scl.ece.ucsb.edu/hoa-waspaa-demo}
\BIBentrySTDinterwordspacing

\end{thebibliography}
%
%
%
%
%
%
%
%
%

\end{sloppy}
\end{document}